# Dilatancy of Shear Transformations in a Colloidal Glass

Y. Z. Lu,[1,*] M. Q. Jiang,[2,3,†] X. Lu,[1] Z. X. Qin,[1] Y. J. Huang,[4,‡] J. Shen,[5]

[1]*School of Materials Science and Engineering, Dalian Jiaotong University, Dalian 116028, People's Republic of China*

[2]*State Key Laboratory of Nonlinear Mechanics, Institute of Mechanics, Chinese Academy of Sciences, Beijing 100190, People's Republic of China*

[3]*School of Engineering Science, University of Chinese Academy of Sciences, Beijing 100049, People's Republic of China*

[4]*School of Materials Science and Engineering, Harbin Institute of Technology, Harbin 150001, People's Republic of China*

[5]*School of Materials Science and Engineering, Tongji University, Shanghai 201804, People's Republic of China*

Shear transformations, as fundamental rearrangement events operating in local regions, hold the key of plastic flow of amorphous solids. Despite their importance, the dynamic features of shear transformations are far from clear. Here, we use a colloidal glass under shear as the prototype to directly observe the shear transformation events in real space. By tracing the colloidal particle rearrangements, we quantitatively determine two basic properties of shear transformations: local shear strain and dilatation (or free volume). It is revealed that the local free volume undergoes a significantly temporary increase prior to shear transformations, eventually leading to a jump of local shear strain. We clearly demonstrate that shear transformations have no memory of the initial free volume of local regions. Instead, their emergence strongly depends on the dilatancy ability of these regions, i.e., the dynamic creation of free volume. More specifically, the particles processing the high dilatancy ability directly participate in subsequent shear transformations. These results experimentally support the Argon's statement about the dilatancy nature of shear transformations, and also shed insight into the structural origin of amorphous plasticity.

Understanding plasticity, i.e., how solids flow, is a classical problem, but still remains open. The plasticity of crystals has been well described in terms of dislocation mobility, deformation twining, grain boundary diffusion, etc. These classical pictures however break down in the face of amorphous solids lacking long-ranger period order. Instead, it is generally recognized that amorphous plasticity results from the accumulation of local irreversible rearrangements occurring within "zones" few to hundreds of atoms/particles wide. Such rearrangement events were originally defined as "shear transformations (STs)" by Argon in 1979 [1] that operate validly in both amorphous alloys and colloidal glasses.

In past decades, great efforts have been made to capture and theorize the ST events. Some landmark works must be highlighted. The atomistic scenario of STs was firstly observed by Falk and Langer in 1998 based on molecular-dynamics simulations [2]. Further they have developed the shear-transformation-zone (STZ) theory [2-4] of amorphous plasticity, where the population of two-state STZs as order parameters enter the flow equation. Associating STs with relaxations on a potential energy landscape, Johnson and Samwer [5] have proposed a cooperative shear model, describing how isolated STs confined within the elastic matrix develop into macroscopic plastic yielding [6-7]. Such a physical picture has been confirmed by atomistic simulations [8-10], continuum modeling [11-12] and direct colloidal-glass visualization [13-15]. As a result, fundamental properties, i.e., activation barriers [16-17], characteristic sizes [7,18] and transformation fields [10,19], of STs can be identified quantitatively.

It has been broadly accepted that STs are transient in time, giving rise spatially to Eshelby fields around them. They do not pre-exist in a glassy structure, but are consequence of complex thermally-activated structural rearrangements driven by an applied shear stress. Nevertheless, this is not suggest that the STs are independent of the structures. For example, it is widely believed that STs occur preferably at sites of high free volume [1,20-21]. Recently, a number of physical parameters have been proposed to predict the structural origin of STs, including low-frequency vibrational mode [22-23], local fivefold symmetry [24], flow unit[25], local yield stress [26], local potential energy [27], flexibility volume [28], etc. These parameters hold more or less predictive power, but to a great extent, from a spatiotemporally statistical view. Building a deterministic one-to-one correlation between dynamics STs and static structures still remains a formidable challenge.

It is not the purpose of the present Letter to directly explore the structural birth of STs. In contrast, we focus on the dynamic process of STs themselves due to their transient nature. The STs dynamics should provide an integrated picture of STs, from their birth to finality. Furthermore, it is interesting to trace the dynamic evolution of free volume during STs. A long-standing idea is that free volume, as a basic structural indicator, is of paramount importance in promoting STs [1,20-21]. But recent works indicate that there seems no strong correlation between particles having high free volume and those undergoing STs [26,29]. Here we attempt to offer an experimental judgement for this disagreement. More importantly, the free volume dynamics during STs could reflect another fundamental property of STs, i.e., the dilatancy. As stated originally by Argon [1], the net effect of the STs is an excess dilatation that must create at least temporarily additional excess free volume. Recently, based on the interaction of STs and STs-mediated free volume dynamics, a constitutive theory of amorphous plasticity is proposed [12]. It indeed predicts that, for glasses with typical strain softening feature, STs always produce free volume, indicating their dilatancy nature. To our best knowledge, however, there has been no experimental proof of the Argon's statement.

In this Letter, we employ a high-speed confocal microscopy to directly track the motion of individual particles in a 3D colloidal glass that is subject to a shear-loading and -unloading cycle. We accurately identify where and when the ST events take place by visualizing the spatiotemporal evolution of local shear strain of particles. Both static free volume prior to STs and its subsequent evolution within STs are also examined. It demonstrates that the emerging STs are weakly correlated with initial free volume, even through the latter is spatially averaged over the second-neighbor shell, a optimum length. Nevertheless, the activations of STs prefer these regions with the high dilatancy capability to create sufficient free volume. These results may increase the understanding of the interplay of STs and free volume, suggesting the dilatancy nature of STs, or in other words, the catalytic effect of dynamic (not static) free volume in STs.

We use 1.55 μm diameter silica particles with a polydispersity smaller than 3.5 % to prepare a colloidal glass [30]. A shear-loading and -unloading cycle is applied to the colloidal glass. The experimental apparatus is the same as that described by Jensen et al [15]. As illustrated in Fig. 1(a), we carefully introduce a transmission electron microscope (TEM) grid deep into the glass. Through a hollow post, we use a piezoelectric translation stage to move the grid to apply a shear and subsequent reverse-shear along the $y$ direction at a very small rate of about $4 \times 10^{-5}$ s$^{-1}$. In the $z$ direction, a high-speed confocal microscopy is utilized to visualize individual particles in a $77 \times 77 \times 27$ μm$^3$ observation volume that is far away from the boundaries. Thus the particles positions in 3D space can be determined. We track the trajectories of individual particles for the 4200 s duration of each cycle by acquiring 3D image stacks every 200 s. Each image stack takes 200 s.

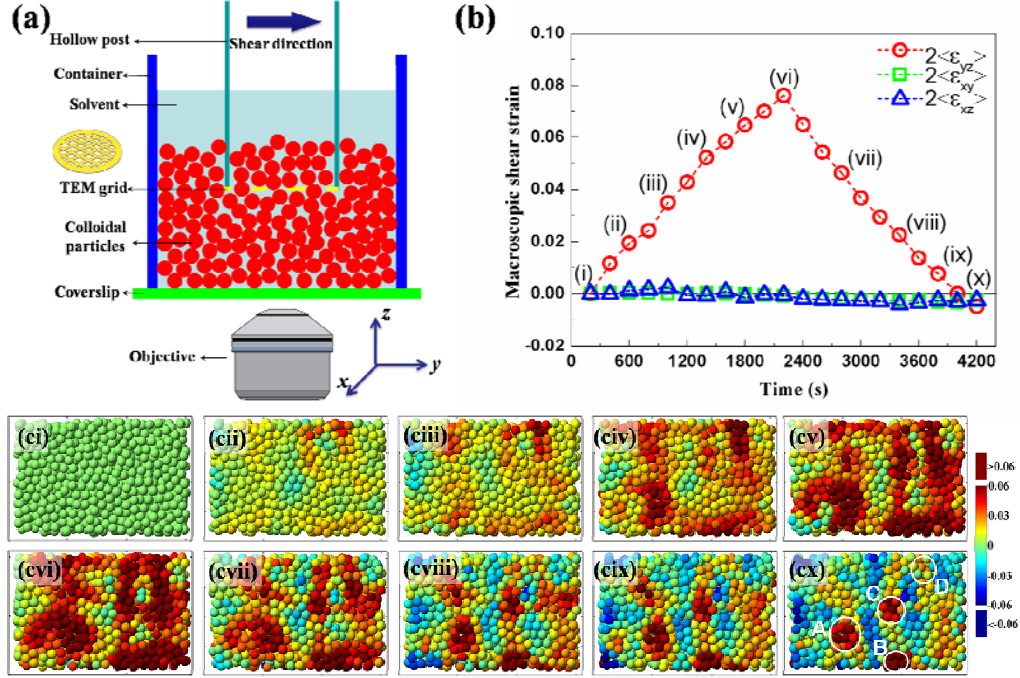

FIG. 1 A colloidal glass under a shear-loading and -unloading cycle. (a) Schematic showing the experimental set-up. (b) Macroscopic shear strain applied on the colloidal glass. (ci)-(cx) Cumulative distribution of local shear strain $\varepsilon_{yz}$ of individual particles in 6-μm-thick $y$-$z$ sections centered at $x$=24 μm, corresponding to the times (i) to (x), respectively, marked in (b).

In order to identify where and when STs occur in this sheared glass, we examine the local shear strain $\varepsilon_{ij}$ of each particle [30-31]. The macroscopic shear strain $2\langle\varepsilon_{ij}\rangle$ is the double average of $\varepsilon_{ij}$ for all particles, whose change during the shear cycle is shown in Fig. 1(b). As we designed that, the contributions to the macroscopic shear strain come mainly from the component $2\langle\varepsilon_{yz}\rangle$, and other components $2\langle\varepsilon_{xy}\rangle$ and $2\langle\varepsilon_{zx}\rangle$ are negligibly small. The $2\langle\varepsilon_{yz}\rangle$ increases linearly from zero to a maximum of about 0.08 at $4\times10^{-5}$ s$^{-1}$, and then return to zero at the same rate. Thus, the local shear strain $\varepsilon_{yz}$ is adopted to monitor STs in the present colloidal glass.

The spatial distribution of the cumulative $\varepsilon_{yz}$ values with the reference zero-strain state in 6-μm-thick $y$-$z$ sections centered at $x$=24 μm is shown in Figs. 1(ci)-(cx), corresponding to the times (i) to (x) marked in Fig. 1(b). With the shear loading, an increasing number of particles participate in shear deformation, but with a spatially inhomogeneous feature. Some localized regions with large $\varepsilon_{yz}$ are emerging, and eventually coalesce at the end of loading. But in fact, not all of these localized regions experience ST events. Here, the shear strain consists of elastic (reversible) and inelastic/plastic (irreversible) parts. Only STs contribute to plastic part, although they are elastic coupling [19,29]. To exactly ferret out the localized regions where STs take place, it is necessary to release the elastic part of shear strain and just retain the plastic part. To this end, we slowly reverse-shear on the glass back to zero macroscopic strain. After the elastic-unloading, most particles return to their initial positions with a very low strain state. We can readily indentify three survived local regions, marked by circles A, B and C, respectively, in Fig. 1(cx), that have relatively high plastic strain. It is reasonably believed that these localized regions experience irreversible STs during the shear deformation.

Next, we explore when the ST events take place in these localized regions. The ST is essentially a local cluster of atoms that undergoes an inelastic shear distortion from one relatively low energy configuration to a second such configuration. This huge local structural modification will accommodate a great change in the local shear strain. Thus, we can use local strain mutation to probe the occurrence of ST events. We then calculate the evolution of the average shear strain $\varepsilon_{yz}$ of the regions A, B and C from initial time to shear-ending (2200 s). As presented in Fig. 2(a), the average shear strain $\varepsilon_{yz}$ of the region A slowly increases during the initial 1000 s. After that, a large strain jump $\Delta\varepsilon_{yz}$, marked by a green line, occurs during 1000 s - 1200 s, indicating the operation of a ST event (labeled as ST-A hereafter). This ST-A event should be underpinned by a significant structural-rearrangement that can be indicated by the variation in the number of nearest neighbors of each particle in the region A [31]. We re-construct typical 3D pictures of particles positions of the region A at 800 s, 1000 s and 1200 s, respectively, as shown in the insets of Fig. 2(a). The red solid spheres in these insets represent those particles that lose three or more nearest neighbors with the reference zero-strain state. Other particles are shown transparently to improve visibility. Comparing the re-constructed 3D pictures at 1000 s and 1200 s, the ST-A event indeed involves a large change of neighbors, corresponding to a pronounced rearrangement of particles. Similar behaviors, including both strain jumps and structural rearrangements, are also observed in the regions B and C, where the ST events (labeled as ST-B and ST-C) occur during 1400 s-1600 s and 1600 s-1800 s, respectively.

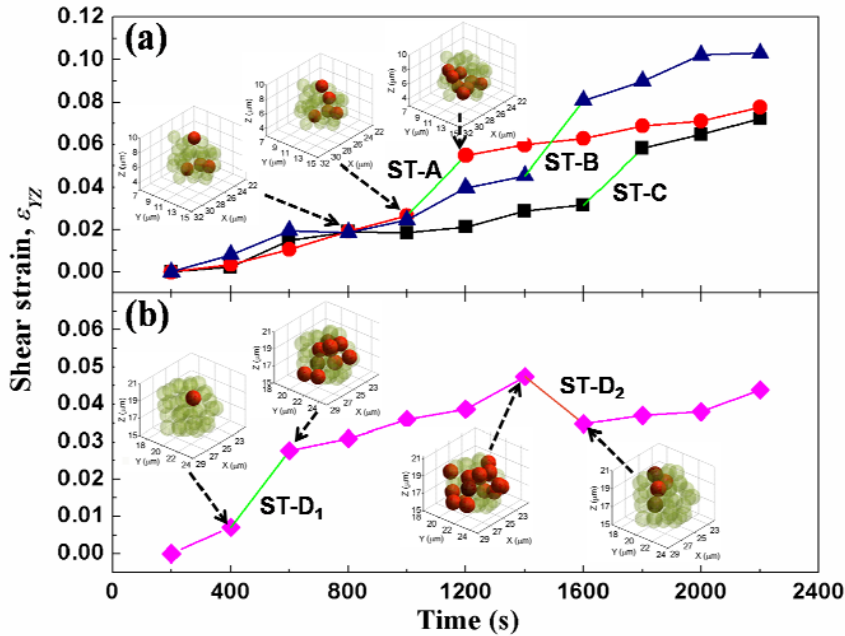

FIG. 2. Shear strain evolution in local regions with shear transformations label as (a) ST-A, ST-B and ST-C, and (b) ST-D from initial time to 2200 s. Insets show the 3D re-constructed pictures of structural rearrangements accompanied with shear transformations.

Falk and Langer [2,32] have proposed that potential ST regions are two-state systems. In the presence of a shear stress, STs can operate along either shear-positive or -negative direction. Obviously, the regions A, B and C only undergo positive ST events that contribute to a sudden increase of local plastic strain (Fig. 2(a)). It is expected that, the negative ST event, analogous to shear-assisted relaxation, will eliminate the local plastic strain. A local ST region may show relatively low plastic strain, if both positive and negative STs occur within it. Therefore, such ST regions may be hidden by their local plastic strain. To uncover them, we carefully examine the shear strain evolution of every region in the glass section (Fig. 1(cx)) from initial time to 2200 s. We indeed find a hidden ST region D, marked by the circle in Fig. 1(cx), where two ST events (labeled as ST-$D_1$ and ST-$D_2$, respectively) with opposite directions take place. The evolution of the average shear strain $\varepsilon_{yz}$ of this region is presented in Fig. 2(b). The ST-$D_1$ event, indicated by the first large strain jump (the green line), occurs during 400 s - 600 s. The ST-$D_2$ event, indicated by the second large strain jump (the orange line), occurs during 1400 s - 1600 s. It is found that the ST-$D_1$-induced strain partly recovers due to the activation of ST-$D_2$. Again, both ST-$D_1$ and ST-$D_2$ correspond to the significant rearrangements of particles in region D, as indicated by the 3D re-constructed pictures shown in the insets of Fig. 2(b).

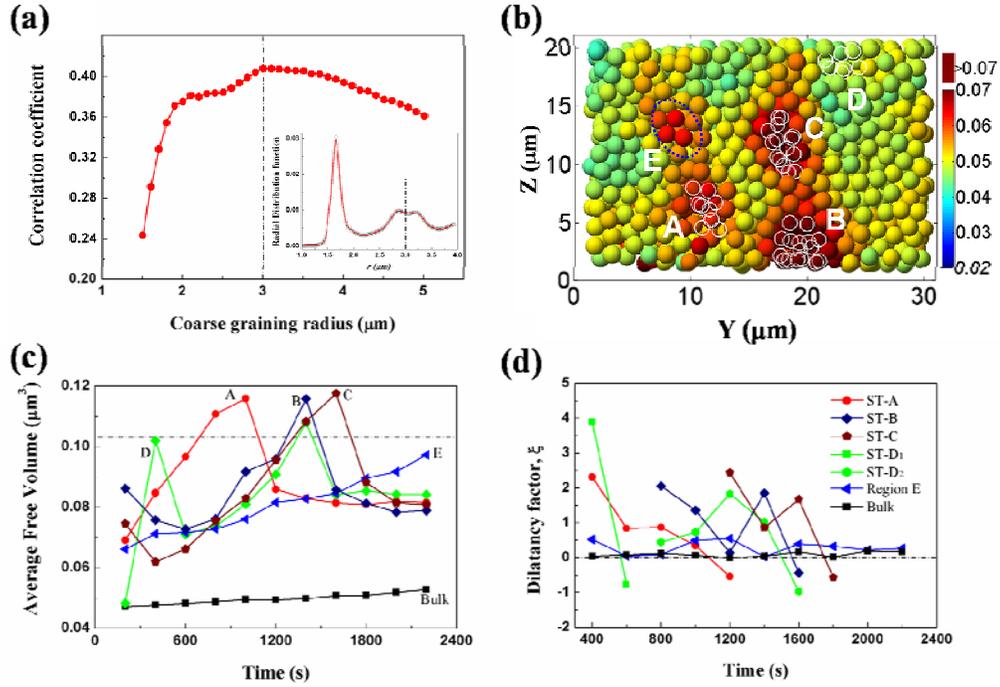

FIG. 3. Correlation between shear transformations and free volumes. (a) Normalized correlation coefficient between local shear strains of particles participated in shear transformations and their initial free volumes as a function of the coarse-graining radius. Inset shows the radial distribution function of the colloidal glass. (b) Spatial distribution of the initial free volume after the optimum coarse-graining in the same section of Fig. 1(c). Small white circles denote the particles participated in shear transformations. (c) Dynamic evolution of average free volumes and (d) dilatancy factors in typical local regions A-E in (b) and the entire bulk glass from initial time to 2200 s.

To explore the dilatancy property of STs, we determine the particle free volumes $v_f$ [30-31] that is believed to be induced by the shear-dilatation effect [33-34]. We quantify the STs-free-volume correlation by defining the normalized correlation coefficient,

$$C_{\varepsilon_{yz}, v_f} = \frac{\sum_k \left( \varepsilon_{yz,k} - \langle \varepsilon_{yz} \rangle \right) \left( v_{f,k} - \langle v_f \rangle \right)}{\sqrt{\sum_k \left( \varepsilon_{yz,k} - \langle \varepsilon_{yz} \rangle \right)^2 \sum_k \left( v_{f,k} - \langle v_f \rangle \right)^2}}, \tag{1}$$

where $k$ denotes a particle that participates in the ST events, $\varepsilon_{yz,k}$ is the local shear strain of such a particle $k$ at 4200 s, $v_{f,k}$ is its initial free volume, and $\langle \varepsilon_{yz} \rangle$ and $\langle v_f \rangle$ are their average values of all particles involved in STs, respectively. According to Fig. 2, the positive strain jumps $\Delta\varepsilon_{yz}$ for all ST events are greater than a critical $\Delta\varepsilon_{yz}^* \sim 0.021$. Therefore, the particles that undergo STs can be identified as those with $\Delta\varepsilon_{yz} \geq \Delta\varepsilon_{yz}^*$ during one time step (200 s). The calculated $C_{\varepsilon_{yz}, v_f}$ is only about 0.24, implying a weak link between the initial free volume and the subsequent ST events. Since STs usually occur within a length scale beyond the nearest-neighbor shell, the coarse-grained free volume may show a better correlation with STs. Here we coarse-grain the initial $v_{f,k}$ by spatially averaging over a particle and its surrounding particles lying within a coarse-graining radius $d$. As shown Fig. 3(a), with increasing $d$, the calculated $C_{\varepsilon_{yz}, v_f}$ increases dramatically from initial 0.24 to a peak of 0.41, and then decreases gradually. It is interesting to find that the optimum $d \approx 3$ μm corresponding to the $C_{\varepsilon_{yz}, v_f}$ peak is very close to the second-neighbor shell of the present colloidal glass. This is confirmed by the RDF shown in the inset of Fig. 3(a). Although the $C_{\varepsilon_{yz}, v_f}$ is improved, the coarse-grained initial free volume still does not show a strong correlation with the ST events. This can be further confirmed from Fig. 3(b) that presents the distribution of the initial free volume after the optimum coarse-graining in the same section of Fig. 1(c). The particles ($\Delta\varepsilon_{yz} \geq \Delta\varepsilon_{yz}^*$) that experience ST-A, ST-B, ST-C and ST-D events are superimposed as marked by small white circles. We indeed find that not all ST events take place in the high free volume regions. In contrast to the ST-A, ST-B and ST-C events at high free volume regions, the ST-D region has a very low-level free volume. Furthermore, the region E with relatively high free volume does not undergo ST events, establishing that the high static free volume is not a necessary precondition for the emergence of STs.

We further examine the free volume dynamics associated with STs. We calculate the evolution of average free volume in the four localized regions (A-D) with STs and the one (E) without STs from initial to 2200 s. The average free volume of the entirely bulk glass is also considered for comparison. All results are shown in Fig. 3(c). The averaged bulk free volume increases slightly during the whole shear deformation. In the region A, the local free volume first increases from initial 0.069 $\mu m^3$ to a maximum of ~0.116 $\mu m^3$ due to the shear-induced dilatation. Immediately after that, the ST-A event occurs and the accompanied significant rearrangement leads to an instantaneous losing of ~27% of free volume. Eventually, the free volume tends towards a steady-state value of ~0.082 $\mu m^3$ due to a dynamic balance between dilatation and relaxation. In the regions B and C, the relatively higher free volume first shows a decrease via relaxation, then followed by an increase to a peak before the occurrence of ST event, rather than a direct increase like that in the region A. This implies that the STs do not seem to have a memory of their initial free volume. Again, after the STs, a steady-state free volume is achieved in the region either B or C. In the region D, where two ST events take place, the initial free volume is small, almost equal to the averaged bulk value. However, such a small initial free volume does not hinder the following ST events. Due to the significant dilatation effect, the local free volume of this region rapidly increases to a very large value of about 0.1 $\mu m^3$, which fertilizes the first ST-$D_1$ event. Accompanied with the ST-$D_1$, the free volume abruptly decrease to a smaller value. Afterwards, the free volume re-increases to the 0.11μm$^3$ level, resulting in the second ST-D$_2$ event, and eventually the free volume decreases towards a steady-state value. From the dynamic evolution of local free volume in the regions A, B, C and D, we notice that a significant free volume creation to a critical value $v_f^*$ is necessary for the activation of STs. Interestingly, it seems that this critical $v_f^*$ has no direct link with the initial value. We calculate the $v_f^*$ by averaging the maximum free volume of all ST particles just before ST events take place. Then we get the critical free volume $v_f^*$ for STs in our colloidal glass is about 0.103 μm$^3$, signed by a dot dash line in Fig.3(c). Obviously in the region E, although its initial value is relatively large, the free volume is always lower than the $v_f^*$ during the whole shear process. Thus, no ST events occur in this region.

It can be seen from Fig. 3(c) that, the main difference in the free volume evolution among regions with and without STs is the dynamic creation of free volume. In the regions A-D with STs, the local free volume can rapidly increase upon the critical $v_f^*$ for preparing the ST operations. But in the region E without STs, the free volumes increases slowly and never reaches the critical $v_f^*$. These results indicate that the creation ability of free volume is pivotal for controlling the activation of STs. In order to measure such a creation ability, we define a dilatancy factor for each particle with a volume of $v_0$:

$$\xi = \frac{\Delta v_f / v_0}{\left| \Delta \varepsilon_{yz} \right|}, \tag{2}$$

which evaluates the increase $\Delta v_f$ of free volume induced by a shear strain increment $\Delta \varepsilon_{yz}$ during one time step. Figure 3(d) gives the calculated results about the evolution of average dilatancy factors in the regions A-E and the bulk glass. Clearly, for these STs regions A-D, the average $\xi$ are all very large before the operation of ST events, and usually higher than 1. Very interestingly, Argon defined a similar ST dilatancy $\bar{\beta}$ and also found $\bar{\beta} = 1$ for shear amorphous soap bubble rafts [35-36]. The region E without STs only shows a relatively low $\xi$ with the maximum of about 0.35 over the whole period. Since the STs highly localize into only the four regions, the average bulk $\xi$ also keeps a low level less than 0.2. Thus, we have a conclusion that the ST events preferentially occur in those regions with the high dilatancy ability of creating free volume, rather than those with the initial high free volume. To prove this more directly, we construct a glass section corresponding to that of Fig. 1(cx) in Fig. 4. The particles with the free volume $v_f^m$ larger than 0.09μm³ are presented as solids spheres. Other particles are displayed transparently to improve visibility. Here, $v_f^m$ is the highest free volume of the particle during the whole shear process during initial to 2200 s. Particle color denotes the dilatancy factor $\xi$ of the particle that is averaged over three adjacent time steps (600 s) prior to reaching the largest free volume $v_f^m$. The particles that undergoes the ST events are again marked by small white circles. We find that there are many particles that have processed large free volume at certain shear times. However, not all of them experience ST events; instead, only those having the high dilatancy factor contribute to STs. An excellent correlation can be observed between them, experimentally supporting that the dilatancy is an inherent nature of ST events in glasses.

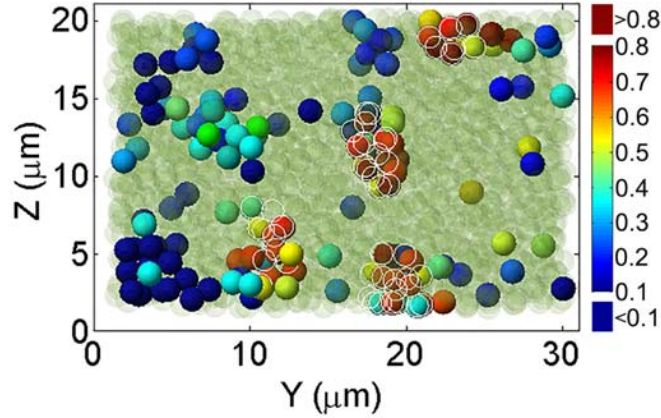

FIG. 4. Spatial distribution of dilatancy factors of particles with high free volume (solid spheres), superimposed upon particles participated in shear transformations (white circles).

To conclude, we perform an in situ observation of the dynamic ST events in a 3D colloidal glass under a simple shear deformation. By following the spatiotemporal evolution of local shear strain of individual particles, we accurately identify where and when the ST events occur in the sheared colloidal glass. Furthermore, we probe the free volume dynamics during STs, and experimentally find that STs do not occur in all local regions with high initial free volume. It implies that STs have no memory of the initial free volume of local regions. By defining the dilatancy factor of particles, we demonstrate that STs prefer these local regions having high dilatancy ability of creating sufficient free volume, where their initial, static free volume is not necessary high. Our results clearly substantiate the idea that the dilatancy is an inherent nature of transient STs, supporting the Argon's statement [1]. Although its precise picture is not very clear, the dilatancy of STs might result from the complex interaction between the ST regions and their neighborhood, closely depending on local free volume, temperature and stress fields.

This work was supported by the National Natural Science Foundation of China (Grant Nos. 11522221, 51401041, 11372315, 51671042, 51671070 and 51274151), the

Strategic Priority Research Program of the Chinese Academy of Sciences (Grant No.XDB22040303), the China Postdoctoral Science Foundation (Grant No. 2015M570242), and the Basic Research Program of the Key Lab in Liaoning Province Educational Department (Grant No. LZ2015011). The first experiments leading to this work were made during Y.Z. Lu's stay in F. Spaepen's and D.A. Weitz's laboratories at Harvard University. He thanks Katharine E. Jensen and J. Zsolt Terdik for their advice and assistance.

*yunzhuohit@gmail.com

[†]mqjiang@imech.ac.cn

[‡]yjhuang@hit.edu.cn